\documentstyle[12pt]{article}
\begin{document}

\title {Sub-critical Closed String Field Theory in D Less Than 26}
\author {Michio Kaku}
\date {Physics Dept., City College of the CUNY, N.Y., N.Y. 10031}
\maketitle

\begin {center}
\Large {\bf Abstract}
\end {center}

In this paper, we construct the second quantized
action for sub-critical closed string field theory
with zero cosmological constant in dimensions
$ 2 \leq D < 26$, generalizing the non-polynomial closed string
field theory action proposed by the author and the Kyoto and MIT
groups for $D = 26$. The proof of gauge invariance is considerably
complicated by the presence of the Liouville field $\phi$ and
the non-polynomial nature of the action. However, we
explicitly show that the polyhedral vertex functions
obey BRST invariance to all orders.
By point splitting methods, we calculate the anomaly contribution
due to the Liouville field, and show in detail that it cancels
only if $D - 26 + 1 + 3 Q ^ 2 = 0 $, in both the bosonized and
unbosonized polyhedral vertex functions.
We also show explicitly that the
four point function generated by this action reproduces the
shifted Shapiro-Virasoro amplitude found from $c=1$
matrix models and Liouville theory in two dimensions.
This calculation is non-trivial because the conformal
transformation from the $z$ to the $\rho$ plane
requires rather complicated third elliptic integrals
and is hence much more involved than the ones found in the
usual polynomial theories.

\section {Introduction}

At present matrix models [1-3] give us a simple and powerful technique for
constructing the S-matrix of two dimensional string theory.
However,
all string degrees of freedom are missing, and hence many of the
successes of the theory are intuitively difficult to interpret
in terms of
string degrees of freedom. Features such as the discrete states [4-7]
and the $w(\infty)$ algebra arise in a rather obscure fashion.

By contrast, Liouville theory [8-9]
manifestly includes all string degrees of freedom,
but the theory is notoriously difficult to solve, even for
the free case.

In order to further develop the Liouville approach, we
present the details of a second quantized field theory of
closed strings defined in $ 2 \leq D < 26$ dimensions with
$\mu = 0$.
(See refs. [10-11] for work on c=1 open string field theory.)

There are several advantages to
presenting a second quantized field formulation of
Liouville theory:

\noindent (a) The $c=1$ barrier, which has proved to
be insurmountable for matrix models, is trivially
breached for Liouville theory (although we no longer
expect the model to be exactly solvable beyond
$c=1$)

\noindent (b) In principle, it should be possible to
present a supersymmetric Liouville theory in field theory form,
which is difficult for the matrix models approach.

\noindent (c) For $c = 1$, the
rather mysterious features appearing in matrix models,
which are intuitively difficult to understand,
have a standard field theoretical interpretation.
For example, \lq\lq discrete states" arise naturally as
string degrees of freedom with discrete momenta when
we calculate the physical states of the theory.
In other words, the $\Phi (X, b,c, \phi )$
field
contains three sets of states. Symbolically, we have:
\begin {equation}
| \Phi ( X , b , c , \phi ) \rangle =
| {\rm tachyon}\rangle +
| {\rm discrete} \, {\rm states} \rangle
+
| {\rm BRST} \, {\rm trivial} \, {\rm states} \rangle
\end {equation}

Also, the structure constants of
$w(\infty)$ arise as the coefficients of the
three-string vertex function, analogous to the
situation in Yang-Mills theory.
We see that $w(\infty)$ is just a subalgebra of the
full string field theory gauge algebra.
For example, if $\langle j , m |$ labels the
$SU(2)$ quantum numbers of the discrete states, then
we can show that the three-string vertex function
$\langle \Phi ^ 3 \rangle $,
taken on discrete states,
reproduces the
structure constants of $w(\infty)$:

\begin {eqnarray}
\langle j _ 1 , m _ 1  | \langle j _ 2 , m _ 2 |
\langle j _ 3 , m _ 3 | V _ 3 \rangle
& \sim &
\Big \langle \Psi  _ { j _ 1 , m_ 1 } ( 0 )
\Psi  _ { j _ 2 , m _ 2 } ( 1 )
\Psi _ { j _ 3 , m _ 3 } ( \infty) \Big \rangle
\nonumber \\
& \sim & ( j _ 1 m _ 2 - j _ 2 m _ 1 ) \delta _ { j _ 3 ,
j _ 1 + j _ 2 - 1 } \delta _ { m _ 3 ,
m _ 1 + m _ 2 }
\end {eqnarray}
where we have made a conformal transformation from the three-string
world sheet to the complex plane, and where the charges
$Q  _ { j,m} = \oint { dz \over 2 \pi i } \Psi _ { j,m} ( z ) $
generate the standard $w(\infty)$ algebra:

\begin {equation}
[ Q _ { j _ 1 , m _ 1 } , Q _ { j _ 2 , m _ 2 } ]
=
( j _ 1 m _ 2 - j _ 2 m _ 1 ) Q  _ { j _ 1 + j _ 2 + 1 ,
m _ 1 + m _ 2 }
\end {equation}

To construct the string field theory action for non-critical
strings, we first begin with the
non-polynomial closed string
action of the 26 dimensional string theory,
first written down by
the author [12] and the Kyoto and MIT groups [13,14]:

\begin {equation}
{\cal L} = \langle \Phi | Q | \Phi \rangle +
\sum _ { n = 3 } ^ \infty \alpha _ { n}
\langle \Phi ^ n \rangle
\end {equation}
where $Q = Q _ 0 ( b _ 0 - \bar b _ 0 )$, $Q _ 0$
is the usual BRST operator, and
where the field $\Phi$ transforms as:

\begin {equation}
\delta | \Phi \rangle =
| Q \Lambda \rangle +
\sum _ { n = 1 } ^ \infty \beta _ { n }  | \Phi ^ n \Lambda \rangle
\end {equation}
where $n$ labels the number of faces of the polyhedra, and there
are more than one distinct polyhedra at each level.
For example, there are 2 polyhedra at $N=6$,
5 polyhedra at $N=7$,
and 14 polyhedra at $N=8$ [12].

If we insert $\delta | \Phi \rangle$ into the action, we find that
the result does not vanish, unless:

\begin {equation}
\label{eq:6}
( -1 ) ^ n \langle \Phi || Q \Lambda \rangle +
n \langle Q \Phi || \Phi ^ { n - 1 } \Lambda \rangle
+
\sum _ { p = 1 } ^ { n-2 }
C _ p ^ n \langle \Phi ^ { n-p} ||
\Phi ^ p \Lambda \rangle = 0
\end {equation}
where the double bars mean that when we join two polyhedra,
the common
boundary has circumference
$2 \pi$. The meaning of this equation is rather simple.
The first two terms on the left hand side represent the
action of $\sum _ i Q _ i$ on the vertex function.
Naively, we expect the sum of these two terms to vanish.
However, naive BRST invariance is broken by the third
term, which has an important interpretation.
This third term consists of polyhedra with rather
special parameters, i.e they are polyhedra
which are at the endpoints of the modular region.
Thus, these polyhedra are actually composites; they
can be split in half, into two smaller
polyhedra, such that the boundary of contact is $2 \pi$. This
is the meaning of the double bars.

(This action also has additional quantum corrections because the
measure of integration $D \Phi (X)$ is not gauge invariant.
These quantum corrections can be explicitly solved in terms of
a recursion relation. These corrections can be computed either by
calculating these loop corrections to the measure [15], or by using the
BV quantization method [16].)

If strings have equal parametrization length $2 \pi$, then we
must triangulate moduli space with cylinders of equal circumference but
arbitrary extension, independent of the dimension of space-time.
Thus, the
triangulation of moduli space on Riemann surfaces remains the
same in any dimension $D$. Therefore, the basic
structure of the action remains the same for sub-critical strings
with equal parametrization length.

What is different, of course, is that the string degrees of freedom
have changed drastically, and a Liouville field $\phi$ must
be introduced.
The addition of the Liouville theory complicates the proof of
gauge invariance considerably, however, since this field
must be inserted at curvature singularities
within the vertex functions, i.e. at the corners of the polyhedra.
This means that the standard proof of gauge invariance formally
breaks down, and hence must be
redone.

This raises a problem, since the
explicit cancellation of these anomalies has only been performed
for polynomial string field theory actions, not the non-polynomial
one. In particular, the anomaly cancellation of the Witten field theory
depends crucially on knowledge of the specific numerical
value of the Neumann functions.
However, the Neumann functions of the non-polynomial field theory are
only defined formally. Explicit forms for them are not known.
Thus, it appears that the cancellation of anomalies seems impossible.

However, we will use point splitting methods, pioneered in [17-19],
which have the advantage that we can isolate those points
on the world sheet where these
insertion operators must be placed, and hence only need to
calculate the anomaly
at these insertion points. Thus, we do not need to have
an explicit form for the Neumann functions; we need only
certain identities which these Neumann functions obey.
The great advantage of the point splitting method, therefore, is
that we can show BRST invariance to all orders in
polyhedra, without having to have explicit expressions for the
Neumann functions.
As an added check, we will calculate the anomaly in two
ways, using both bosonized and unbosonized ghost variables.

Thus, we will first calculate the anomaly contribution,
isolating the potential divergences coming from the
insertion points and show that they sum to zero.
Then we will show that our theory
reproduces the standard shifted Shapiro-Virasoro amplitude.

\section {BRST Invariance of Vertices}

We will specify our conventions by introducing a field
which combines the string variable $X ^ i $ (where $i$ labels
the Lorentz index) and the
Liouville field $\phi$. We introduce
$\phi ^ { \mu } $ where $\mu = 0, 1  , 2 , ... D$ and
where $\phi ^ { D }$ corresponds to the
Liouville field, so that $\phi ^ \mu = \{ X ^ i , \phi \}$.

The first quantized action is given by:

\begin {equation}
S = { 1 \over 8\pi } \int d ^ 2 \xi \sqrt { \hat g }
\left \{
g ^ { ab } \left ( \partial _ a X ^ i \partial _ b X _ i
+ \partial _ a \phi \partial _ b \phi \right )
+ Q \hat R \right \}
\end {equation}

The holomorphic part of the
energy-momentum tensor is therefore:
\begin {eqnarray}
\label{eq:T}
T _ { zz } ^ \phi & = &
- { 1 \over 2 } \left( \partial _ z \phi ^ \mu \right ) ^ 2
- { Q _ \mu  \over 2 } \left ( \partial _ z ^ 2
\phi ^ \mu \right )
\nonumber \\
T _ { zz } ^ {\rm gh } & = &
{ 1 \over 2 } \left( \partial _ z \sigma \right ) ^ 2
+ { 3 \over 2 }
\left (\partial _ z \sigma ^ 2 \right )
\end {eqnarray}
where we have bosonized the ghost fields via $c = e ^ \sigma$
and $b = e  ^ { - \sigma }$ and where $ Q ^ \mu = ( 0 , Q )$.
Demanding that the central charge of the
Virasoro algebra vanish implies that:
\begin {equation}
[ L _ n , L _ m ] =
( n - m ) L _ { n+m}
+ { c \over 12 }
n ( n ^ 2 - 1 ) \delta _ { n+m, 0 }
\end {equation}
with total central charge:
\begin {equation}
c = D + 1 + 3 Q ^ 2 - 2 6 = 0
\end {equation}
so that $Q = 2 \sqrt 2$ for $D = 1$ (or for two dimensions
if we promote $\phi$ to a dimension).
Notice that the
ghost field has a background charge of $-3$ and
the $\phi ^ \mu$ field has a background charge of
$ Q ^ \mu = ( 0 ,  Q ) $.
This allows us to collectively place the bosonized
ghost field and the $\phi ^ \mu$ field together
into one field.
We will use the index $M$ when referring to the
collective combination of the string variable, the
Liouville field, and the bosonized field.
We will define

\begin {eqnarray}
Q ^ M & = & \{ 0 , Q , -3 \}
\nonumber \\
\phi ^ M & = & \{ X ^ i , \phi , \sigma \}
\end {eqnarray}

To calculate the insertion factors in the vertex function, we
must analyze the terms in the first quantized action proportional
to the background charge:

\begin {equation}
{ Q ^ M  \over 8 \pi } \int {\sqrt {g} } R \, \phi ^ M d ^ 2 \xi
\end {equation}
where we have normalized the curvature on the world sheet such
that $\int { \sqrt {g} } R d ^ 2 \xi = 4 \pi \chi $ where $\chi$ is the
Euler number.
In general, the curvature on the string world sheet is zero,
except at isolated points where the strings join.
At these interior points, the curvature is a delta function,
such that $\int \sqrt {g} R d ^ 2 \xi = - 4 \pi$ around these points.
This means that $N$-point vertex functions, in general, must have
insertions proportional to:

\begin {equation}
\prod _ { j = 1 } ^ { 2(N-2)}
\left( e ^ { -  Q ^ M \phi ^ M / 2 } \right) _ j
\end {equation}
where $j$ labels the $2(N-2)$ sites where we have curvature
singularities on the string world sheet.
These insertions, in fact, are the principle complication facing us
in calculating the anomalies of the various vertex functions.

The vertex
is then defined as:

\begin {equation}
| V _ N \rangle = \int B _ N | V _ N \rangle _ 0
\end {equation}
where $B _ N$ consist of line integrals of
$b$ operators defined over Beltrami differentials
(see the Appendix for conventions for the vertex function)
and $| V _ N \rangle _ 0$ is the standard vertex function
given as an over-lap condition on the string and ghost degrees of
freedom which must satisfy the usual BRST condition:

\begin {equation}
\sum _ { i = 1 } ^ N Q _ i | V _ N \rangle _ 0 = 0
\end {equation}

Notice that it is the presence of this factor $B _ N$ which
prevents the vertex function from being trivially BRST
invariant.
The reason for this is that $B _N$ contains
line integrals of the $b$ operators, defined over Beltrami differentials
$\mu _ k$,
such that:

\begin {equation}
T _ { \mu _ k } = \{ Q , b _ {\mu _ k}\}
\end {equation}

Whenever $Q$ is commuted past a term in $B _N$, it creates an
expansion or contraction of some of the modular parameters
within the polyhedral vertex function. The deformation generated
by $T _ { \mu _ k }$
is given
as a total derivative in the modular parameter $\tau _ k$, i.e.

\begin {equation}
\int d \tau _ k T _ { \mu _ k } \sim
\int d \tau _ k { \partial \over \partial \tau _ k }
\end {equation}

When this
deformation is integrated
over the modular parameter, we find only the endpoints of the
modular region. However, the endpoints of the modular region
are where the polyhedra splits into two smaller polyhedra, connected
by a common boundary of $2 \pi$.
This, in turn, reproduces the residual terms $\langle \Phi ^ { n-p}
|| \Phi ^ p \Lambda \rangle$ appearing
in
eq. (\ref{eq:6})
which violate naive BRST invariance.
Thus, the importance of this $B_N$ term is that it
gives the corrections to the naive BRST invariance equations.

Fortunately, the factor $B _ N$ remains the same even for
the sub-critical case independent of the dimension of space-time.
Therefore, we can ignore this term and shall concentrate instead on
the properties of $| V _ N \rangle _ 0$, which is defined as:

\begin {eqnarray}
|V _ N \rangle _ 0 & = &
\left (
\prod _ { j = 1 } ^ { 2 ( N-2) } e ^ { - ( Q ^ M \phi ^ M / 2 ) _ j } \right )
 \int
\delta ( \sum _ { i = 1 } ^ N
p _ i ^ M + Q ^ M ) \prod _ { i = 1 } ^ N P_ i
\nonumber \\
& \times &
{\rm exp } \,
\Bigg \{
\sum _  { r,s } ^ N \sum _ { n,m = 0}
 ^ \infty
{ 1 \over 2 }
N _ { nm } ^ { rs } \alpha  _ { -n } ^ { M r }
\alpha _ { -m } ^ { M  s }
\Bigg \}
\nonumber \\
& \times &
{\rm exp } \,
\Bigg \{
\sum _  { r,s } ^ N \sum _ { n,m = 0}
 ^ \infty
{ 1 \over 2 }
N _ { nm } ^ { rs } \tilde \alpha  _ { -n } ^ { M r }
\tilde \alpha _ { -m } ^ { M  s }
\Bigg \}
\left ( \prod _ { i = 1 } ^ N d ^ M p _ i | p _ i ^ M \rangle
\right)
\end{eqnarray}
where $P _ i $ represents the operator which rotates the string
field by $2 \pi$,
where $j$ labels the insertion points, where we have deliberately
dropped an uninteresting constant,
and where the
state vector $| p _ i ^M \rangle $ and
the Neumann functions are defined in the Appendix.

For our calculation, we would like to commute the insertion
operator directly into the vertex function. Performing the
commutation, we find (for $N=3$):
\begin {eqnarray}
| V _ 3 \rangle _ 0 & = &
\int
\delta ( p _ 1 ^ M + p _ 2 ^ M + p _ 3 ^ M + Q ^ M )
\prod _ { i=1} ^ 3 P _ i
\, {\rm exp } \,
\Bigg \{
\sum _  { r,s } ^ 3 \sum _ { n,m = 0}
 ^ \infty
{ 1 \over 2 }
N _ { nm } ^ { rs } \alpha  _ { -n } ^ { M r }
\alpha _ { -m } ^ { N  s }
\nonumber \\
& - & { 1 \over 3 } { \bar Q ^ M \over 2 }
\Big [ \sum _ { r = 1 } ^ 3
\sum _ { n=1 } ^ \infty
{ ( -1 ) ^ n
\over 2n } \alpha _ { - 2n } ^ { M r }
+ \sum _ { r,s } ^ 3 N _ {00} ^ { rs }
\alpha _ 0 ^ { M r }
\nonumber \\
& + &
\sum _ { r,s } ^ 3 \sum _ { n = 1 } ^ \infty
N _ { n0 } ^ { rs }
\alpha _ { - n } ^ { M r }
- \sum _ { n = 1 } ^ \infty \sum _ { r,s } ^ 3
N _ { n m } ^ { rs }
\alpha _ { -n } ^ { Mr } \cos { m \pi/ 2 }
\Big ]
\Bigg \}
\left ( \prod _ { i =1 } ^ 3
d ^ M p _ i | p _ i \rangle \right)
\end {eqnarray}
where $\bar Q ^ M = \{ 0 , - i Q , -3 \}$.
The factor $1/3$ appearing before the background charge arises because
we have broken up the insertion operator into three equal pieces, each defined
in terms of the three different harmonic
oscillators.
(For simplicity, we have only presented the holomorphic part of
the vertex function, and deleted $\tilde \alpha$ operators for
convenience. It is understood that all vertex functions contain
both the $\alpha$ and $\tilde \alpha$ operators.)

With these conventions, we now wish to show that the
vertices of the  non-polynomial theory are
BRST covariant. For the three-string vertex, this means:
$ \sum _ { i = 1 } ^ 3 Q _ i | V _ 3 \rangle _ 0 = 0$.

Naively, this calculation appears to be trivial, since
the vertex function
simply represents a delta function across three overlapping strings.
Hence, we expect that the three contributions to $Q$ cancel exactly.
However, this calculation is actually
rather delicate, since there are potentially
anomalous contributions at the joining points.

Previous calculations of this identity were limited by the
fact that they used specific information about the three-string
vertex function. We would like to use a more general method
which will apply for the arbitrary $N$-string vertex function.
The most general method uses point-splitting.

We wish to construct a conformal map from the multi-sheeted,
three-string world sheet configuration
in the $\rho$-plane to the flat, complex $z$-plane.
Fortunately, this map was constructed in
[12]:

\begin {equation}
\label {eq:map}
{ d \rho (  z )
\over d z }
= C { \prod _ { i = 1 } ^ { N - 2 }
\sqrt  { ( z - z _ i ) ( z -  \tilde z _ i ) }
\over
\prod _ { i = 1 } ^ N  ( z - \gamma _ i ) }
\end {equation}
where the $N$ variables
$\gamma _ i$ map to
points at infinity (the external
lines in the $\rho$ plane) and the $N-2$ pair of variables
$( z _ i, \tilde z _ i ) $ map to the
points where two strings collide, creating the
$i$th vertex (which are interior points in
the $\rho$ plane).

The set of complex numbers
$\{ C , z _ i ,  \tilde z _ i , \gamma _ i \} $
constitute an initial set of $2 + 4( N-2) + 2N = 6N - 6 $ unknowns.
In order to achieve the correct counting, we must impose a number
of constraints. First, we must set the length of the
external strings at infinity to be $\pm \pi$. In the
limit where $z \rightarrow \gamma _ i$, we have:

\begin {equation}
{\rm lim } \, \, { d \rho ( z ) \over dz } \rightarrow
{ \pm 1 \over z - \gamma _ i }
\end {equation}
This gives us $2N$ constraints. However, by projective invariance
we have the freedom of selecting
three of the $\gamma _ i$ to be $\{ 0 , 1 , \infty \}$
Then we must subtract two, because of
over-all charge conservation (taking into account that
there are charges due to the Riemann cuts as well as charges
located at $\gamma _i$.)
Thus we have
$2N + 6 -2 = 2N + 4$ constraints.

Next, we must impose the fact that the overlap of two colliding strings
at
the $i$th vertex is given by $ \pi$, such that the interaction
takes place instantly in proper time $\tau$.
This is gives us:

\begin {equation}
\pm i \pi  = \rho ( z _ i ) - \rho ( \tilde z _ i )
\end {equation}

This gives us $2(N-2)$ additional constraints, for a
total of $4N$ constraints.
Thus, the number of variables minus the number of constraints is given by
$2N - 6$.
But this is precisely the number of Koba-Nielsen variables
necessary to describe $N$ string scattering, or the number
of moduli necessary to describe a Riemann surface with $N$
punctures consisting of cylinders of equal circumference and
arbitrary extension.

These moduli can be described in terms of the proper time
$\tau$ separating the $i$th and $j$th vertices, as well as the
angle $\theta$ separating them.

We can define
\begin {equation}
\hat \tau _ { ij } = \tau _ { ij } + i \theta _ { ij }
=
\rho ( z _ i ) - \rho ( z _ j )
\end {equation}
where $\tau _ { ij}$ is the proper time separating
the $i$th and $j$th vertices, and $\theta _ { ij}$ is
the relative angle between them.

There are precisely $2N - 6$ independent variables contained
within the $\hat \tau _ { ij }$, as expected. (Not all
the $\hat \tau _ { ij }$ are independent.)

In summary, we find that:

\begin {eqnarray}
\{ C , z _ i , \tilde z _ i , \gamma _ i\} & \rightarrow &
6N-6 \, {\rm unknowns}
\nonumber \\
\{ \rho ' ( \gamma _ i ) , \rho ( z _ i) - \rho ( \tilde z _ i )
\} & \rightarrow & 4N \, {\rm constraints}
\nonumber \\
\{ \tau _ { ij } + i \theta _ { ij } \} & \rightarrow &
2N - 6 \, {\rm moduli}
\end {eqnarray}

The conformal map, with these constraints, describes
$N$ point scattering consisting of three-string vertices only.
This is not enough to cover all of moduli space. In addition,
we find a \lq\lq missing region" [20]. For example, we must include
the $2N - 6$ moduli necessary to describe the lengths of the sides of
an $N$ sided polyhedra. The moduli describing the various polyhedra
are specified by setting $\tau _ { ij }$ all equal to each other.
In other words, on the world sheet, the polyhedral interaction
takes place instantly in $\tau$ space.
Then the $2N-6$ variables necessary to describe the polyhedra can
be found among the $\theta _ { ij }$.

Now that we have specified the conformal map,
we can begin the calculation of the BRST invariance of
the vertex functions.

First, we will find it convenient to transform the
BRST operator $Q$ into a line integral over the $\rho$ plane.
For the three-point vertex function, we have three line integrals
which, for the most part, cancel each other out (because of the
continuity equations across the vertex function). The only
terms which do not vanish are the ones which
encircle the joining points $z _ i$ and $\tilde z _ i$.

Written as a line integral, the BRST condition becomes:

\begin {eqnarray}
\sum _ { i = 1} ^ 3 Q _ i | V _ 3 \rangle _ 0  & = &
\oint _ { C _ 1 + C _ 2 + C _ 3 }
{ d \rho \over 2\pi } c ( \rho )
\nonumber\\
& \times &
\left \{
- { 1 \over 2 }
( \partial _ z \phi ^ \mu ) ^ 2
+ { d c \over d\rho } b ( \rho )
+ { Q \over 2 } ( \partial _ z \phi ^ \mu ) ^ 2
\right \} | V _ 3 \rangle _ 0
\end {eqnarray}
where $C _ i$ are infinitesimal curves which
together comprise circles which go around
$\rho ( z _ i)$ and $\rho (\tilde z _ i )$.
Notice that this expression is, strictly speaking, divergent
because they are defined at the joining point $ z _ i$, where
these quantities, in general, diverge.
To isolate the anomaly, we will now make a conformal transformation
from the $\rho$ plane to the $z$ plane.
When two operators are defined at the same point, as in
$(\partial _ z\phi ^ \mu ) ^ 2 $, we will point split them
by introducing another variable $z '$ which is infinitesimally
close to $z$.
Then our expression becomes:

\begin {eqnarray}
\sum _ { i = 1 } ^ 3 Q _ i | V _ 3 \rangle _ 0
& =&
\oint _ { C _ 1 +C _ 2 + C _ 3 }
{dz \over 2 \pi i }
c( z )
\nonumber \\
& \times & \Bigg \{
- { 1 \over 2 } \left ( { dz ' \over  dz } \right)
\partial \phi _ \mu ( z ' ) \partial \phi ^ \mu ( z )
\nonumber \\
&+&
\left ( { dz ' \over dz } \right ) ^ 2
{ dc \over dz } b ( z ' )
+ { Q \over 2 }
\partial ^ 2 \phi ( z ) \Bigg  \}
| V _ 3 \Big \rangle _ 0
\end {eqnarray}
where $z'$ is infinitesimally close to $z$,
where $\mu$ ranges over the $D$ dimensional string modes as well
as the $\phi$ mode,
where $b$ and $c$ are the usual reparametrization ghosts,
and the $C _ i$ are now infinitesimally small
curves in the $z$-plane which encircle
the joining point, which we call $z _ 0$.
In making the transition from the $\rho$ plane to the
$z$ plane, we have made the
re-definition:

\begin {equation}
c ( \rho ) =  { d \rho \over d z }
c ( z ) ; \quad
b ( \rho ) =
\left ( { d \rho \over d z }  \right ) ^ { -2 }
b  ( z )
\end {equation}

The major complication to this calculation
is that the Liouville $\phi$ field does
not transform as a scalar. Instead, it transforms as:

\begin {equation}
\label{eq:phi}
\phi ( \rho ) \rightarrow \phi (  z ) +
{ Q \over 2 } \log \Big | { d z \over d \rho } \Big |
\end {equation}

This means that the energy-momentum tensor $T$ transforms as:

\begin {eqnarray}
T _ { \rho  \rho } &  \rightarrow &
\left ( { d z \over d \rho } \right) ^ 2 T _ { zz }
+ \left( { Q \over 2 } \right ) ^ 2 S
\nonumber \\
S & = &
{ z ''' \over z '}   - { 3 \over 2 } \left ( { z '' \over z  ' } \right ) ^ 2
\end {eqnarray}
where $S$ is called the Schwartzian.
The form of the Schwartzian that is most crucial for our
discussion will be:

\begin {equation}
\left ( { Q \over 2 } \right ) ^ 2 S =
T _ { zz } ^ \phi \left ( \partial _ z \phi \rightarrow
{ Q \over 2 } \partial _ z \log | { dz \over d \rho } | \right)
\end {equation}

This complicates the calculation considerably, since it means that there
are
subtle insertion factors
located at delta-function
curvature singularities in the vertex function.
These add non-trivial $\phi$
contributions to the calculation.

\section {Point Splitting}

In order to perform this sensitive calculation, we will use the
method of point splitting.

Let us examine the behavior of the various variables near the
splitting point $z _ 0$ using the original conformal map in
eq.(\ref{eq:map}).
Near this point, we have:

\begin {eqnarray}
\label {eq:join}
{ d \rho \over dz } &= &
a { \sqrt { z - z _ 0 } } +
b { \sqrt { z - z _ 0 } } ^ 3 +  ...
\nonumber \\
\rho ( z ) & = &
\rho ( z _ 0 ) +
{ 2 \over 3 } a ( z - z _ 0 ) ^ { 3/2 }
\left( 1 + { 3 \over 5 }
{ b \over a }
( z - z _ 0 ) + ... \right)
\end {eqnarray}

Now let us define $\epsilon = z - z _ 0$ and power expand
these functions for small $\epsilon$.
For the purpose of point splitting, we introduce
the
variable $z '$, which is infinitesimally close to both
$z$ and $z _ 0$, and is defined implicitly through the
equation:

\begin {equation}
\label {eq:a}
\rho ( z ' ) = \rho ( z ) + { 2 \over 3 } a \delta
\end {equation}
where $\delta $ is an infinitesimally small constant, which we will
later set equal to zero.

We will find it convenient to introduce the following function:

\begin {equation}
f ( \epsilon ) =
z ' - z _ 0 =
\epsilon \left\{ 1 + \sum _ { n=1 } ^ \infty
f _ n (\epsilon ) \delta ^ n \right\}
\end {equation}

We can easily solve for the coefficients $f _ n$ by power
expanding the following equation:

\begin {eqnarray}
\rho ( z ' ) - \rho ( z ) & = &
{ 2 \over 3 } a \delta
\nonumber \\
& = &
{ 2 \over 3 } a \left (
f ^ { 3/2 } ( \epsilon )  - \epsilon ^ { 3/2 } \right )
+
{ 2 b \over 5 }
\left ( f ^ { 5/2 }
( \epsilon ) - \epsilon ^ { 5/2 }
\right )
+ ...
\end {eqnarray}

By equating the coefficients of $\delta$, we find:

\begin {eqnarray}
f _ 1 & = &
{ 2 \over 3 } \epsilon ^ { - 3/2 }
\left ( 1 - p \epsilon \right) + ...
\nonumber \\
f _ 2 & = &
- { 1 \over 9 } \epsilon ^ { -3 }
+ ...
\nonumber \\
f _ 3 & = &
{ 4 \over 81 } \epsilon ^ { - 9/2 }
\left ( 1 - p \epsilon \right )
+ ...
\end {eqnarray}
where $p = b/a$.

In our calculation, we will find potentially divergent quantities, such
as $ 1 / ( z ' - z ) $ and $ dz' / dz$, so we will
power expand all these quantities in terms of $f _ n$ in a double
power expansion in $\epsilon$ and $\delta$.

Then we easily find:

\begin {eqnarray}
{ 1 \over z ' - z } & = &
{ 1 \over f ( \epsilon )  - \epsilon  }
\nonumber \\
& = & { 1 \over \epsilon f _ 1 \delta }
\left ( 1 - { f _ 2 \over f _ 1 } \delta
+ \delta ^ 2 \left(
{ f _ 2 ^ 2 \over f _ 1 ^ 2 }
- { f _ 3 \over f _ 1 } \right )
+ ... \right )
\nonumber \\
{ 1 \over ( z '  - z ) ^ 2 } & = &
{ 1 \over \epsilon ^ 2 f _ 1 ^ 2 \delta ^ 2 }
\Bigg [ 1 - 2 \delta { f _ 2 \over f _ 1 }
\nonumber \\
& + &
 \delta ^ 2 \left ( - 2 { f _ 3 \over f _ 1 } +
3 { f _ 2 ^ 2 \over f _ 1 ^ 2}  \right ) + ...
\Bigg ]
\end {eqnarray}

Also:

\begin {eqnarray}
{ d z ' \over dz } & = &
{ df( \epsilon )  \over dz }
\nonumber \\
& = &
1 + \sum _ { n = 1 } ^ \infty
\delta ^ n ( \epsilon f _ n ) '
\end {eqnarray}

We also find:
\begin {eqnarray}
{ d z ' \over d z }
{ 1 \over ( z ' - z ) ^ 2 }
& = &
\epsilon ^ { -2 }
f _ 1 ^ { -2 }
\Bigg [
( \epsilon f _ 2 ) ' - 2 f _ 3 f _ 1 '
\nonumber \\
& +  &  3 f _ 2 ^ 2 f _ 1 ^ { -2 }
+ ( \epsilon f _ 1 ) ' ( -2 f _ 2 f _ 1 ^ { -1 } )
+
... \Bigg]
\nonumber \\
& = &
\epsilon ^ { -2 } \left ( { 5 \over 48 } +
{ p \epsilon \over 12 } \right)
+ ...
\end {eqnarray}

\begin {eqnarray}
\left ( { dz ' \over d z } \right ) ^ 2
{ 1 \over ( z ' - z  ) ^ 2 }
& = &
\epsilon ^ { -2 } f _ 1 ^ { -2 }
\Bigg [
- 4 ( \epsilon f _ 1 ) ' f _ 2
f _ 1 ^ { -1 }
\nonumber \\
& + &
( \epsilon f _ 1 ) ' + 2 ( \epsilon f _ 2 ) '
\nonumber \\
&-&
 2 f _ 3 f _ 1 ^ { -1 } + 3 f _ 2 ^ 2
f _ 1 ^ { -2 } + ...
\Bigg
]
\nonumber \\
& =&
\epsilon ^ { -2
}
\left ( { 29 \over 48 }
+ { 13 \over 12 }
p \epsilon \right )
+ ...
\end {eqnarray}

\begin {eqnarray}
\left (
{ d z ' \over d z } \right) ^ 2
{ 1 \over z ' - z }
& = &
\epsilon ^ { -1 }
f _ 1 ^ { -1 }
\left ( -f _ 2 f _ 1 ^ { -1 }
+
2 ( \epsilon f _ 1 ) '
\right )
\nonumber \\
& = &
- { 3 \over 4 }
\epsilon ^ { -1 }
+ ...
\end {eqnarray}
(The terms contained in $ ... $ correspond to terms which can be discarded
in our approximation.
For example, we will take the limit as
$\epsilon \rightarrow 0$ first, and then take the limit as
$\delta \rightarrow 0$. This allows us to eliminate
powers of $\delta$ occurring with negative exponent.
Also, there is a Riemann cut in the map in eq. (\ref{eq:map}),
so we will choose the regularization scheme in ref. [19].
We can do this by altering eq. (\ref{eq:a}) slightly.
We can define our point splitting by re-expressing our
operators in terms of
two new variables, $z _1$ and $z _2$, such that
$\rho ( z _ 1 ) = \rho ( z ) + (2/3) a \delta$
and $\rho ( z _ 2 ) = \rho ( z ) - ( 2/3 ) a\delta$.
Then operators are defined in terms of averaging over
$z _1 $ and $z _ 2$.
This averaging corresponds to
choosing $\rho ( z ' ) = \rho  ( z ) + ( 2/3 ) a \delta$
and then discarding odd powers of $\delta$.)

Now that we have defined all our regularized expressions,
we can begin the process of calculating the anomaly.
Let us first analyze the anomaly coming from the term
$\partial _ { z ' } \phi ( z ' ) \partial \phi ( z )$.
We will commute this expression into the Neumann functions.
We will then extract from this a
$c$-number expression which represents the anomaly.

When we shove this term into the vertex function,
we pick up quantities which look like $n m N _ { nm } ^ { rs }
\omega _ r ^ n \tilde \omega _ s ^ m$, where $ \omega  = e ^ \zeta$,
where, following Mandelstam, we take $\zeta$ to be
a local variable defined on the closed string, such that
$\zeta = \tau + i \sigma$. $\zeta$ and $\rho$ coincide for
the closed string lying on the real axis.
Fortunately, we know how to calculate
this term in terms of $z$ variables.

Let us differentiate the expression in eq. (\ref{eq:N})
in the Appendix:

\begin {eqnarray}
\label{eq:41}
{ d \over d \zeta _ s } N ( \rho _ r , \tilde \rho _ s )
& = &
\delta _ { rs }
\left \{ { 1 \over 2 }
\sum _ { n \geq 1 } \omega _ r ^ { -n }
( \tilde \omega _ s ^ n + \tilde \omega _ s ^ { * n } )
+ 1 \right \}
\nonumber \\
& + &
{ 1 \over 2 }
\sum _ { n , m  \geq  0 }
n N _ { nm } ^ { rs }
\omega _ r ^ n ( \tilde
\omega _ s ^ m + \tilde
\omega _ s ^ { * m } )
\nonumber \\
& = &
{ 1\over 2 } { d z \over d \zeta _ r }
\left ( { 1 \over z - \tilde z } +
{ 1 \over z - \tilde z ^ * } \right )
\end {eqnarray}
and (by letting $\tilde z _ s $ go to $\gamma _ s$):

\begin {equation}
\label{eq:42}
\delta _ { rs } +
\sum _ { n\geq 1 }
n N _ { n0 } ^ { rs } \omega _ r ^ n =
{ d z \over d \zeta _ r }
{ 1 \over z - \gamma _ s }
\end {equation}

A double differentiation leads to:

\begin {eqnarray}
{ d \over d \zeta _ r }
{ d \over d  \tilde \zeta _ s }
N ( \rho _ r , \tilde \rho _ s )
& = &
\delta _ { rs } { 1 \over 2 }
\sum _ { n \geq 1 } n \omega _ r ^ { -n }
\tilde \omega _ s ^ n +
\nonumber \\
& + &
{ 1 \over 2 } \sum _ { n,m \geq 1 }
nm N _ { nm } ^ { rs }
\omega _ r ^ s \tilde \omega _ s ^ m
\nonumber \\
& = &
{ 1 \over 2 }
{ d z _ r \over d \zeta _ r }
{ d \tilde z _ s \over d \tilde \zeta _ s }
{ 1 \over ( z - \tilde z ) ^ 2 }
\end {eqnarray}

We will now perform the calculation in two ways, using
unbosonized ghost variables $b$ and $c$, and then using
the bosonized ghost variable $\sigma$.

\subsection {Method I: Unbosonized Ghosts}

With these identities, it is now an easy matter to calculate
the action of the BRST operator on the vertex function in terms
of unbosonized ghost variables $b$ and $c$.
Let the brackets $ \langle$ $\rangle$ represent
the $c$-number expression what we obtain when we perform this
commutation. Then we can show that the background-independent
terms yield:

\begin {eqnarray}
\Big \langle \partial _ z ' \phi ^ \mu ( z ' ) \partial
_ z \phi ( z ) ^ \nu \Big \rangle & =&
- { 1 \over ( z ' - z ) ^ 2 }  \delta ^ { \mu \nu }
\nonumber \\
\Big \langle c ( z ) b ( z ' )  \Big \rangle
& = &
- { 1 \over z ' - z }
\end {eqnarray}

With these expressions,
we can now calculate the contribution to the anomaly
due to $\partial _ {  z ' } \phi ( z ' ) \partial _ z
\phi ( z )$ and $ \partial _ z c ( z ) b ( z ' )$.
This calculation is simplified because the ghost insertion
factor disappears in the $b-c$ formalism.

We find (dropping the background-dependent terms, for the moment):
\begin {eqnarray}
\label {eq:bc}
\sum _ { i = 1 } ^ 3 Q _ i | V _ 3  \rangle _ 0 & = &
\oint _ { C _ 1 + C _ 2 + C _ 3 }
{ dz \over 4 \pi i }
\Bigg \{
c (z ) \Big [
- { d z ' \over dz }
\Big \langle \partial _ { z ' } \phi ^ \mu ( z ' )
\partial _ z \phi ^ \mu ( z ) \Big \rangle
\nonumber \\
& + &
2 \left ( { d z ' \over d z } \right ) ^ 2
\Big \langle { dc \over dz } ( z ) b ( z ' ) \Big \rangle
\Big ]
- 2 { d c \over dz }
\left ( { dz ' \over d z } \right ) ^ 2
\Big \langle c ( z )
b( z ' ) \Big \rangle
\Bigg \} | V _ 3 \rangle _ 0 + ...
\nonumber \\
& = &
-  \oint _ { C _ 1 + C _ 2 + C _ 3 }
{ dz \over 4 \pi i }
\Bigg \{
c (z ) \Big [
- { d z ' \over dz }
{ 1 \over ( z ' - z ) ^ 2 }
\nonumber \\
& + &
2 \left ( { d z ' \over d z } \right ) ^ 2
\partial _ z { 1 \over z ' - z }
\Big ]
- 2
{ d c \over dz }
\left ( { d z ' \over dz } \right ) ^ 2
{ 1 \over z ' - z }
\Bigg \} | V _ 3 \rangle _ 0
+ ...
\nonumber \\
& = &
- \oint _ { C _ 1 + C _ 2 + C _ 3 }
{ dz \over 4 \pi i }
\Bigg \{
c ( z )
\Big [
- \epsilon ^ { -2 } \left (
{ 5 \over 48 } + { p \epsilon \over 12 } \right )
\nonumber \\
& + &
2 \epsilon ^ { -2 }
\left ( { 29 \over 48 }
+ { 13 \over 12 } p \epsilon \right )
\Big ]
-2
{ d c \over dz }
\left (
- { 3 \over 4 } \epsilon ^ { -1 } \right )
\Bigg \}
| V _ 3 \rangle _ 0
+ ...
\end {eqnarray}
where $...$ are terms which are
background-dependent.
Now let us combine the three arcs $C _i$ into one circle
which goes around the joining point $z _ 0$.
Integrating by parts, we find:

\begin {eqnarray}
\label {eq:46}
\sum _ { i = 1 } ^ 3 Q _ i | V _ 3 \rangle _ 0 & = &
\oint _ { z _ 0 } { d z\over 2 \pi i }
\Bigg \{  { p c ( z ) \over z - z _ 0 }
\left [ { D \over 24 } + { 1 \over 24 }
- { 13 \over 24 }
\right ]
\nonumber \\
& + & { d c ( z ) \over dz }
{ 1 \over z - z _ 0 }
\left [
{ 5 D \over 96 }
+
{ 5 \over 96 }
- { 65 \over 48 } \right ]
\Bigg \}
| V _ 3 \rangle _ 0
+ ...
\end {eqnarray}

The last part of the calculation  is perhaps the most crucial,
i.e. calculating the contribution of the term
$\partial _ z \phi $ to the anomaly which are background-dependent.
Normally, this term does not contribute at all. However,
in the presence of the insertion operator at the joining points
$z _ i $ and $ \tilde z _ i$,
this term does in fact contribute an important part to the
anomaly.

Our task is to shove the operator $\partial _ z \phi $ into
the vertex function and calculate terms proportional to the
background charge $Q$.
We find:

\begin {eqnarray}
\partial _ \rho  \phi _ r ( \rho )  | V _ 3 \rangle _ 0
\nonumber \\
& = &
( -i ) ^ 2 { Q \over 2 } \Bigg ( \sum _ { n,m \geq 0 } \sum _ r
\omega _ r  ^ n
N _ { nm } ^ { rs } \cos ( m \pi  / 2 )
\nonumber \\
&& - \sum _ { n \geq 1 } \sum _ r \omega _ r ^ n
N _ { n0 } ^ { rs }
\Bigg ) | V _ 3 \rangle _ 0 + ...
\end {eqnarray}
We immediately recognize the terms on the right as being
functions of $1 / ( z - z _ i ) $ and
$ 1 / ( z - \gamma _ i )$ in eqs. (\ref{eq:41}) and
(\ref{eq:42}) for the case $r \neq s$.

The contribution of the anomaly from the
insertion operator is therefore given by:
\begin {eqnarray}
\Big \langle \partial _ \rho \phi (\rho ) \Big \rangle
&=&
- { Q \over 2 }
\left [
{ 1 \over 2 }
\sum _ { i = 1 } ^ { M-2 }
\left ( { 1 \over z - z _ i }
+
{ 1 \over z - \tilde z _ i } \right )
-
\sum _ { i = 1 } ^ M { 1 \over z - \gamma _ i }
\right ]
\nonumber\\
& = &
{ d z \over d \rho }
{ Q \over 2 } \partial _ z \log | { d z \over d\rho } |
\end {eqnarray}

(As an added check on the correctness of this calculation, notice
that the last step reproduces the desired transformation property of
the $\phi$ field in eq. (\ref{eq:phi}),
which has an additional contribution due to
the background charge $Q$. Thus, when we insert this term into
the expression for the energy-momentum tensor, we simply
reproduce the Schwartzian.)

Given this expression, we
can now calculate the contribution of the background-dependent
terms to the anomaly.
This contribution is:

\begin {eqnarray}
\label{eq:44}
... & = &
\oint _ { z _ 0 } { d z \over 2 \pi i }
c ( z )
\left [
- { 1 \over 2 } \Big \langle
\partial _ z \phi \Big \rangle ^ 2
+ { 1 \over 2 }
Q
{ d \rho \over d z }
\partial _ z
\left( { dz \over d \rho }
\Big \langle \partial _ z \phi \Big \rangle \right)
\right ] | V _ 3 \rangle _ 0
\nonumber \\
& = &
\oint _ { z _ 0 } { d z \over 2 \pi i }
c ( z ) { Q ^ 2 \over 4 }
\left (
{ 5 \over 8 } {   1 \over ( z - z _ 0 ) ^ 2 }
+
{ 1 \over 2 } p { 1 \over  z - z _ 0  }
\right ) | V _ 3 \rangle _ 0
\end {eqnarray}

The last step is to put all terms
together. Combining the results of eq. (\ref{eq:46})
and (\ref{eq:44}), we now easily find:

\begin {equation}
\left \{  p c ( z _ 0 ) \left [
{ D \over 24 } - { 13 \over 12 } + { 1 \over 24 }
+ { 1 \over 8 } Q ^ 2 \right ]
+ { d c ( z _ 0 ) \over d z }
\left [ { 5 D \over  96 }
- { 65 \over 48 }
+ { 5 \over 96 }
+ { 5 \over 32 } Q ^ 2 \right ]
\right \} | V _ 3 \rangle _ 0
\end {equation}
which cancels if:

\begin {equation}
D - 26 + 1 + 3 Q ^ 2 = 0
\end {equation}
which is precisely the consistency equation for Liouville theory
in $D$ dimensions.
Thus, the vertex is BRST invariant.

\subsection {Method II: Bosonized Ghosts}

Next, we will show that the calculation can also be performed using
the bozonized ghost variable $\sigma$.
We exploit the fact that the $X$, $\phi$, and $\sigma$
field can be arranged in the
same composite field $\phi ^ M$.

When we commute $\partial _ z \phi ^ M$ into the vertex function,
we find that the $\sigma$ ghost variables contribute an almost
identical contribution as the $\phi$ variable.

Let us redo the calculation in two parts.
We will calculate the background-independent terms
first. This means dropping the $b$ and $c$ terms in
eq. (\ref{eq:bc}) and replacing the $\phi ^ \mu$
field by $\phi ^ M$.
The calculation is straightforward, and yields:

\begin {eqnarray}
\sum _ { i = 1 } ^ 3 Q _ i | V _ 3 \rangle _ 0 & = &
\oint _ { z _ 0 } { d z\over 2 \pi i }
\Bigg \{  { p { e ^ \sigma ( z ) } \over z - z _ 0 }
\left [ { ( D + 2 ) \over 24 }
\right ]
\nonumber \\
& + & { d { e ^ \sigma ( z ) } \over dz }
{ 1 \over z - z _ 0 }
\left [
{ 5 ( D + 2 ) \over 96 }
\right ]
\Bigg \}
| V _ 3 \rangle _ 0
+ ...
\end {eqnarray}

Next, we must calculate the background-dependent terms.
We can generalize the equation which determines how
the fields change when they are commuted past the
insertion operators:

\begin {equation}
\Big \langle \partial _ \rho \phi ^ M (\rho ) \Big \rangle
=
{ d z \over d \rho }
{ Q ^ M \over 2 } \partial _ z \log | { d z \over d\rho } |
\end {equation}

The crucial complication is that
the quadratic term in the energy-momentum tensor in
eq. (\ref{eq:T}) for the
$\phi$ field and the $\sigma$ field differs by a factor
of $-1$.
This means that when we insert this expression into the
BRST operator $Q$, we pick up an extra $-1$ factor, so the
contribution to the anomaly from the background-dependent terms
now becomes:

\begin {eqnarray}
... & = &
\oint _ { z _ 0 } { d z \over 2 \pi i }
e ^ { \sigma ( z ) }
\left [
\pm { 1 \over 2 } \Big \langle
\partial _ z \phi ^ M \Big \rangle ^ 2
+ { 1 \over 2 }
Q ^ M
{ d \rho \over dz }
\partial _ z
\left( { dz \over d \rho }
\Big \langle \partial _ z \phi ^ M \Big \rangle \right)
\right ] | V _ 3 \rangle _ 0
\nonumber \\
& = &
\oint _ { z _ 0 } { d z \over 2 \pi i }
e ^ \sigma { ( z ) }  { Q   ^ 2 - 3 ^ 2 \over 4 }
\left (
{ 5 \over 8 } {  1 \over ( z - z _ 0 ) ^ 2 }
+
{ 1 \over 2 } p { 1 \over z - z _ 0  }
\right ) | V _ 3 \rangle _ 0
\end {eqnarray}
where the $ - (+)$ sign appears with the $\phi ( \sigma)$ operator.

Now, let us put all the terms together in the calculation.
We find:

\begin {equation}
\left \{ p e ^ { \sigma ( z _ 0 ) } \left [
{ D + 2 \over 24 }
+ { 1 \over 8 } ( Q ^ 2 - 3 ^ 2  ) \right ]
+ { d e ^ { \sigma ( z _ 0 ) } \over d z }
\left [ { 5 ( D + 2 ) \over  96 }
+ { 5 \over 32 } ( Q ^ 2 - 3 ^ 2 ) \right ]
\right \} | V _ 3 \rangle _ 0
\end {equation}
Once again, we find that the anomaly cancels if we set:

\begin {equation}
D + 2 + 3 ( Q ^ 2 - 3 ^ 2 )= 0
\end {equation}
as desired.
Thus, the anomaly cancels in both the bosonized and the unbosonized
expressions, although each expression is qualitatively quite
dissimilar from the other. This is a check on the correctness of our
calculations.

Similarly, the anomaly can be cancelled for all
non-polynomial
vertices. For an $N$-sided polyhedral vertex, we
first notice that
the BRST operator $Q$, once it is commuted past
the various $b$ operators,
vanishes on the bare vertex because of the continuity
equations, except at the
$2(N-2)$ joining points $z _i$ and $\tilde z _ i$.

Second, we notice that
the conformal map around
each joining point in eq. (\ref{eq:join}) is
virtually
the same, no matter
how complicated the polyhedral vertex function may be.
All the messy dependence on the various constraints are
buried within $\rho ( z _ 0 )$ and $p$.
Fortunately,  the dependence on these unknown
factors cancels out of the calculation. This is why the calculation
can be generalized to all polyhedral vertices.

Thus, the calculation of the anomaly cancellation can be performed
on each of the various joining points
$z _i$ and $\tilde z _ i$ separately.
But since the calculation is basically the same for each of
these joining points, we have now shown that all possible
polyhedral vertex functions are all anomaly-free.

Notice that this proof does not need to know
the specific value of the Neumann functions.
The entire calculation just depended on knowing
the derivatives of eq. (\ref{eq:N}) and how
various operators behaved when commuted into the
vertex function.

\section {Shifted Shapiro-Virasoro Amplitude}

The next major test of the theory is whether it reproduces the
shifted Shapiro-Virasoro amplitude. This calculation is
highly non-trivial, since the conformal map between the multi-sheeted
string-scattering Riemann sheet to the complex plane is
very involved. Unlike the conformal maps found
in light cone theory, or even the maps found in Witten's
open string theory [21-22], the non-polynomial theory yields very
complicated conformal maps.

Fortunately, for the four-point function, all conformal maps
are known exactly, in terms of elliptic functions, and the
calculation can be performed [23-24].

For the four point function, the map in eq. (\ref{eq:map}) can be integrated
exactly.
We use the identity:

\begin {equation}
{ ( z - z _ 1 ) ( z - \tilde z _ 1 )
( z - z _ 2 ) ( z - \tilde z _ 2 )
\over \prod _ {i = 1 } ^ 4 ( z - \gamma _ i ) }
=
1 + \sum _ { i = 1 } ^ 4
{ A _ i \over z - \gamma _ i }
\end {equation}
where we define $ z _ i = i a _ i + b _ i$ and $\tilde z _ i
= - i a _ i  + b _ i $ for {\it complex} $a _ i$ and $b _ i$, and:

\begin {equation}
A _ i = {  \left [ ( \gamma _ i - b _ 1 ) ^ 2 + a _ 1 ^ 2 \right ]
\left [ ( \gamma _ i - b _ 2 ) ^ 2 + a _ 2 ^ 2  \right ]
\over
\prod _ { j = 1 , j \neq i } ^ 4
( \gamma _ i - \gamma _ j ) }
\end {equation}

Then we can split the integral into two parts, with the result:

\begin {eqnarray}
\rho ( z ) & = &
\rho _ 1 ( z ) + \rho _ 2 ( z )
\nonumber \\
\rho _ 1 ( z ) & = &
\int _ { y _ 1 } ^ y
 { N dz
\over \sqrt { ( z - z _ 1 ) (z  - \tilde z _ 1 )
( z - z _ 2 ) ( z - \tilde z _ 2 ) } }
\nonumber \\
\rho _ 2 ( z ) & = &
\sum _ { i = 1 } ^ 4
\int _ { y _ 1 } ^ y
{ N A _ i dz \over
( z - \gamma _ i )
\sqrt { ( z - z _ 1 )( z - \tilde z _ 1 )( z - z _ 2 )
(z - \tilde z _ 2 ) } }
\end {eqnarray}

Written in this form, we can now perform all integrals exactly,
using third elliptic integrals in eq.(\ref{eq:F}) and
eq. (\ref{eq:pi}) in the Appendix.
It is then easy to show:

\begin {eqnarray}
\rho _ 1 ( z ) & = &
N g  F ( \phi , k ' ) =
N g {\rm tn } ^ { -1 }
\left [ \tan \phi , k
' \right ]
\nonumber \\
\rho _ 2 ( z ) & = &
\sum _ { i = 1 } ^ 4
{ g N A _ i
\over
a _ 1 + b _ 1 g _ 1 - g _ 1 \gamma _ i }
\Bigg (
g _ 1 F ( \phi , k ' )
\nonumber \\
& + &
{ \omega _ i - g _ 1 \over
1 + \omega _ i ^ 2 }
\left [ F ( \phi , k ' ) + \omega _ i ^ 2
\Pi ( \phi , 1 + \omega _ i ^ 2 , k ' )
+
\omega _ i ( \omega _ i ^ 2 + 1 ) f _  i
\right]
\Bigg )
\end {eqnarray}

where:
\begin {eqnarray}
\omega _ i &=& { a _ 1 + b _ 1 g _ 1 - \gamma _ i g _ 1
\over b _ 1 - a _ 1 g _ 1 - \gamma _ i }
\nonumber \\
f _ i &=&
{ 1 \over 2 } ( 1 + \omega _ i ^ 2 ) ^ { - 1 /2 }
( k ^ 2 + \omega _ i ^ 2 ) ^ { - 1/2 }
\nonumber \\
&& \times \ln
{ ( k ^ 2 + \omega _ i ^ 2 ) ^ { 1/2 }
- ( 1 - \omega _ i ^ 2 ) ^ { 1/2 }
{ \rm dn } u
\over
( k ^ 2 + \omega _ i ^ 2 ) ^ { 1/2 }
+ ( 1 + \omega _ i ^ 2 ) ^ { 1/2 }
{ \rm dn } u }
\nonumber \\
\phi & = & {\rm arc tan}
\left( { y - b _ 1 + a _ 1 g _ 1
\over a _ 1 + g _ 1 b _ 1 - g _ 1 y } \right )
\end {eqnarray}
and where:
\begin {eqnarray}
A ^ 2 & = & ( b _ 1 + b _ 2 ) ^ 2 +
( a _ 1 + a _ 2 ) ^ 2
, \, \,  B ^2  =
( b _ 1 - b _ 2 ) ^ 2 + ( a _ 1 - a _ 2 ) ^ 2
\nonumber \\
g _ 1 ^ 2 &= & [4 a _ 1 ^ 2 - ( A - B) ^ 2 ] /
[ ( A + B ) ^ 2 - 4 a _ 1 ^ 2 ]
, \, \,
g  =  2 / ( A + B )
\nonumber \\
y _ 1 & = & b _ 1 - a _ 1 g _ 1
, \, \, { k '} ^ 2 =  1 - k ^ 2 = 4AB / ( A + B ) ^ 2
\nonumber \\
u &=  & {\rm  dn} ^ { -1 } ( 1 - { k '}  ^ 2  {\rm sin} ^ 2 \phi )
\end {eqnarray}

After a certain amount of algebra, this expression simplifies considerably
to:
\begin {eqnarray}
\rho ( z ) & = &
\sum _ { i = 1 } ^ 4
{ g N A _ i \over
a _ 1 + b _ 1 g _ 1 - g _ 1 \gamma _ i }
{ \omega _ i - g _ 1 \over 1 + \omega _ i ^ 2 }
\nonumber \\
&& \times
\left [ \omega _ i ^ 2
\Pi ( \phi , 1 + \omega _ i ^ 2  , k ' )
+ \omega _ i  ( \omega _ i ^ 2 + 1 )
 f _i \right ]
\end {eqnarray}

Now that we have an explicit form for the conformal map from the
flat $z$ plane to the $\rho$ plane, in which string scattering takes place,
we must next
impose the constraint that the
overlap between two colliding strings
is given by $\pi$.
This is satisfied by imposing:

\begin {eqnarray}
\label{eq:delta}
\pi & = &
{\rm Im }
\left[ \rho ( z _ 1 ) - \rho ( y _ 1 ) \right ]
\nonumber \\
& = &
- { \pi \over 2 }
g N
{ ( \omega _ i - g _ 1 ) \omega _ i
\over a _ 1 + b _ 1 g _ 1 - g _ 1 \gamma _i }
\sum _ { i = 1 } ^ 4
{ A _ i \Lambda _ 0 ( \beta _ i , k )
\over
\sqrt { ( 1 + \omega _ i ^ 2 ) ( k ^ 2 +
\omega _ i ^ 2 ) } }
\nonumber \\
& = &
{ - \pi \over 2 }
\sum _ { i = 1 } ^ 4 \alpha  _ i
\Lambda _ 0 ( \beta _ i , k )
\end {eqnarray}
where $\alpha _ i = NA _ i \left[ ( \gamma _ i - b _ 1 ) ^ 2 + a _ 1 ^ 2
\right ] ^ { - 1/2}
\left[ ( \gamma _ i - b _ 2 ) ^ 2 + a _ 2 ^ 2 \right ] ^ { -1/2}$,
where we have used eq. (\ref{eq:lambda}),
where we have set $y = z _ 1$, so that $\tan \phi = i
$,
and where $ \sin ^ 2 \beta _ i = ( 1 + \omega _ i ^ 2 ) ^ { -1 } $.
We have also used the fact that:

\begin {equation}
\Pi ( \phi , 1 + \omega _ i ^ 2 , k ' )
=
 - { 1 \over 2 } \pi i
{ \sqrt { 1 + \omega _ i ^ 2 } \over
\sqrt { k ^ 2 + \omega _ i ^ 2 } }
{ \Lambda _ 0 (\beta _ i , k ) - 1
\over \omega _ i }
\end {equation}

Next, we must calculate
the separation between the two vertices and the relative angle
of rotation between them.
The proper time separating the two interactions is given by:
\begin {eqnarray}
\tau & = &
{\rm Re } \left [ \rho ( z _ 2 ) - \rho ( z _ 1 ) \right ]
\nonumber \\
& = &
g \sum _ { i = 1 } ^ 4  N A _ i
{ \omega _i ^ 2 ( \omega _ i - g _ 1 ) \Pi
( \pi / 2 , 1 + \omega _ i ^ 2, k ' )
\over
( a _ 1 + b _ 1 g _ 1 - g _ 1 \omega _ i )
( 1 + \omega _ i ^ 2 ) }
\nonumber \\
& = &
- K ( k ' ) \sum _ {i=1 } ^ 4
\alpha _ i Z ( \beta _ i , k ' )
\end {eqnarray}
where we have used eq. (\ref{eq:Z})
and the fact that:

\begin {eqnarray}
\Pi ( \phi , 1 + \omega _ i ^ 2 , k ' )
& = &
\Pi ( \phi _ 2 , 1 + \omega _ i ^ 2 , k ' )
- \Pi ( \phi _ 1 , 1 + \omega _ i ^ 2 , k ' )
\nonumber \\
\Pi ( \alpha ^ 2 , k ) & = &
- { \alpha K Z ( {\rm arcsin } \alpha ^ { -1 } , k )
\over
\sqrt { ( \alpha ^ 2 -1 )( \alpha ^ 2 - k ^ 2 ) } }
\end {eqnarray}
and:

\begin {eqnarray}
{\rm tan } \phi _ 1 & = & i , \, \,
\phi _ 1 = i \infty
\nonumber \\
{\rm tan } \phi _ 2 & = &
{ i \over k } , \, \,
\phi _  2 =
{\rm arcsin } { 1 \over k ' }
\end {eqnarray}
which we can show by setting $y = z _ 1 , z _ 2 $.

Now that we have an explicit form for $\tau$, the next
problem is to differentiate it and find the Jacobian
of the transformation
of $\tau$ to $x$.

By differentiating, we find:

\begin {eqnarray}
d \tau & = &
- \sum  _ { i = 1 } ^ 4 \alpha _ i
{ r ^ 2 ( \beta _ i , k ' ) K ( k ' )
- E ( k ' )
\over
r ( \beta _ i , k ' ) }
d \beta _ i
\nonumber \\
& = &
{ \pi \over 2 }
K ( k ) ^ { -1 }
\sum  _ { i = 1 } ^ 4
{ \alpha _ i d \beta _ i
\over r ( \beta _ i , k ' ) }
\nonumber \\
& = &
{ \pi N \over 2gK ( k ) }
\sum _ { i = 1 } ^ 4
{ d \gamma _ i
\over
\prod _ { j=1 , j \neq i }
( \gamma _ i - \gamma _ j ) }
\end {eqnarray}
where $r ( \theta , k ' ) =
\sqrt { 1 - k ^ { \prime 2 } \sin ^ 2 \theta }$
and
where we have used eq. (\ref{eq:deriv}) in the Appendix.
We have also used the
fact that the derivative
of $\pi$ in eq. (\ref{eq:delta}) is a constant, so:

\begin {equation}
0  =
\sum _ { i = 1 } ^ 4 \alpha _ i
{ E(k ) - k ^ { \prime 2 }
\sin ^ 2 \beta _ i K ( k )
\over r ( \beta _ i , k ' ) }
d \beta _ i
\end {equation}

This explicit conformal map allows us to calculate the four-point
amplitude.
We first write the amplitude in the $\rho$ plane, and then make a
conformal map to the $z$-plane.   Let the modular parameter be
$\hat \tau = \tau + i \theta $, where $\tau$ is the distance
between the splitting strings, and $\theta $ is the relative
rotation.
Then, with a fair amount of work,
one can find the Jacobian from $\hat \tau$ to $\hat x$.

Let us define $\hat x$ as:
\begin {equation}
\hat x =
{ ( \gamma _ 2 - \gamma _ 1 ) ( \gamma _ 3 - \gamma _ 4 )
\over ( \gamma _ 2 - \gamma _ 4 )(
\gamma _ 3 - \gamma _ 1 ) }
\end {equation}
so that:
\begin {equation}
d \hat x = \hat x ( 1 - \hat x )
{ ( \gamma _ 1  - \gamma _ 3 ) ( \gamma _ 2
- \gamma _ 4 )
\over
( \gamma _ 1 - \gamma _ 2 ) ( \gamma _ 1 - \gamma _ 3 )(
\gamma _ 1 - \gamma _ 4 ) } d \gamma _ 1
\end {equation}

Putting everything together, we now find:

\begin {equation}
{ d \hat \tau \over d \hat x }
=
- { \pi N \over
2 K( k ) g \hat x ( 1 - \hat x )
( \gamma _ 1 - \gamma _ 3 )( \gamma _ 2 - \gamma _ 4)}
\end {equation}

If we take only the tachyon component of
$|\Phi \rangle$, then the
four point amplitude can be written as:

\begin {eqnarray}
A _ 4 &=& \langle V _ 3 | { b _ 0 \bar b _ 0 \over
L _ 0 + \bar L _ 0 - 2 } | V _ 3 \rangle
\nonumber \\
& = &
\int d \tau d \theta
\, \Big \langle
V ( \infty ) V ( 1 )
\left ( \int _ C d z { dz \over dw } b _ { zz } \right)
\left ( \int _ C { d \bar z \over d \bar w } b_ { \bar z \bar z }
\right )
V ( \hat x ) V ( 0 ) \Big \rangle
\nonumber \\
& = &
\int d ^ 2 \hat \tau
\, \Bigg | {\rm exp } \, \left [
\sum _ { i } \left ( i p _ i \cdot
\phi ( i ) + \epsilon _ i  \phi ( i ) \right ) \right )
A _ G \Bigg | ^ 2
\end {eqnarray}
where we must sum over all permutations so that
we integrate over the entire complex plane,
where $b _ 0$ defined in the $\rho$ plane
transforms into $\int _ C dz (dz/dw) b _ { zz }$ in the
$z$-plane,
where $C$ is the image in the $z$-plane of a
circle in the $\rho$ plane
which slices the intermediate closed string,
where $V ( z ) = c ( z ) \tilde c ( z ) V _ 0 ( z ) $,
where $V _ 0$ is the tachyon vertex without ghosts,
and
where the ghost part $A _ G$ equals:

\begin {eqnarray}
A _ G &=&
\int _ C { d z \over 2 \pi i }
{ d z \over d w }
{\rm exp }
\left \{
- \sum _ { i \leq j }
\langle \sigma _ i \sigma _ j
\rangle +
\sum _ j
\langle \sigma _ j \sigma _ + ( z ) \rangle
\right \}
\nonumber \\
& = &
\int _ C { d z \over 2 \pi i }
{ d z \over dw }
{ \prod _ { i < j } ( \gamma _ i - \gamma _ j )
\over \prod _ { j = 1 } ^ 4
( z - \gamma _ j ) }
\noindent \\
& = &
2 { g \over \pi c }  \hat x ( 1 - \hat x ) K ( k )
( \gamma _ 1 - \gamma _ 3 ) ^ 3 ( \gamma _ 2 - \gamma _ 4 ) ^ 3
\end {eqnarray}
(Notice that we have made a conformal transformation from the
$\rho$ world sheet to the $z$ complex plane. In general, we pick
up a determinant factor, proportional to the
determinant of the Laplacian defined on the world sheet.
However, after making the conformal transformation,
we find that the determinant of the Laplacian
on the flat $z$-plane
reduces to a constant. Thus, we can in general ignore this
determinant factor.)

Putting the Jacobian, the ghost integrand, and the
string integrand together, we finally find:

\begin {equation}
A _ 4 = \int d ^ 2 \hat x \Big |
\hat x ^ { 2 p _ 1 \cdot p _ 2 }
( 1 - \hat x ) ^ { 2 p _ 2 \cdot p _ 3 } \Big | ^ 2
\end {equation}
In two dimensions, we have
$p _ i \cdot p _ j  = p _ i p _ j -
\epsilon _ i \epsilon _ j$ where
$\epsilon _ i = \sqrt {2} + \chi _ i p _ i $,
where $\chi$ is the \lq\lq chirality" of the tachyon
state,
so we reproduce the integral found in matrix models
and Liouville theory. (The amplitude is non-zero only if
the chiralities are all the same except for one external line.)

However, so far the region of integration does {\it not}
cover the entire complex $z$-plane. This is because
we have implicitly assumed in
the constraints $ \hat \tau _ { ij } =
\rho ( z _ i ) - \rho ( z _ j )$
that there is no four-string interaction.
However, as we have shown in [20], the complete
region of integration contains a \lq\lq missing region"
which is precisely filled by the four string interaction.
This calculation carries over, without any change, to the
$D < 26$ case.

With the missing region filled by the four-string tetrahedron graph,
we finally have the complete shifted Shapiro-Virasoro amplitude,
as expected.

Lastly, we would like to mention the direction for possible
future work. Two problems come to mind.
The most glaring deficiency of this approach is that
we have set the cosmological constant to zero.
However, the theory becomes quite non-linear for  non-zero
cosmological constant, so the calculations become much more
difficult.

The second problem is that we have not shown the equivalence of
this approach to the Das-Jevicki action [25-6], which is the second quantized
field theory of matrix models.
This action is based strictly on the tachyon, so we
speculate that, once we gauge away the BRST trivial states and
integrate out the
discrete states, our action should reduce down to the Das-Jevicki action
(for $\mu  = 0$).
This problem is still being investigated.

\section {Acknowledgments}
We would like to acknowledge partial support from
CUNY-FRAP 6-64435 and NSF PHY-9020495.

\section{Appendix}

We will find it convenient to define the holomorphic
expressions for the operators as follows. (It is understood
that we must double the operators in order to describe
the closed string.)
If we define $\phi ^ M = \{ X ^ i , \phi , \sigma \}$,
then:

\begin {equation}
\partial _ z \phi ^ M =
\sum _ { n = - \infty } ^ \infty
\{ - i \alpha _ n  ^ i , - i \phi _ n , \sigma _ n \}
z ^ { - n - 1 }
\end {equation}
where:
\begin {equation}
[ \phi _ n ^ M , \phi _ m ^ N ] =
n \delta ^ { MN  } \delta _ { n,-m}
\end {equation}
where $\delta ^ { MN } = {\rm diag} \, \{ \delta ^ { ij } ,
1 , 1 \} $.

Physical states without ghost indices
are defined via the conditions:

\begin {eqnarray}
L_ n \ | \Phi \rangle & = & \bar L _ n | \Phi \rangle = 0
\nonumber \\
( L _ 0 - 1 ) | \Phi \rangle
& = & ( \bar L _ 0 - 1 ) | \Phi \rangle = 0
\nonumber \\
(L _ 0 - \bar L _ 0 ) | \Phi \rangle & = &
0
\end {eqnarray}

The tachyon state is defined as:
\begin {equation}
| p ^ \mu \rangle = | p ^ i , \epsilon \rangle =
e ^ { i p \cdot X + \epsilon \phi } ( 0 ) | 0 \rangle
\end {equation}
where $\alpha _ 0 ^ \mu | p \rangle = p ^ \mu | p \rangle $.

To solve for $\epsilon$ and the mass of the tachyon,
we must solve the on-shell condition:
\begin {equation}
L _ 0 | p ,\epsilon \rangle =
\bar L _ 0 | p , \epsilon \rangle =
\left ( { 1 \over 2 } p _ i ^ 2 - { 1 \over 2 }
\epsilon ( \epsilon + Q ) \right )
| p , \epsilon \rangle
\end {equation}
so that:

\begin {equation}
p _ i ^ 2 - \epsilon ( \epsilon + Q ) - 2 = 0
\end {equation}
To put this in more familiar mass-shell form, let us
define $E =  \epsilon + ( 1/2 ) Q $.
Thus, the mass-shell condition can be written as:

\begin {equation}
p _ i  ^ 2 - E ^ 2 = - \left ( { 1 \over 4 } Q ^ 2 - 1 \right ) ^ 2
= - m ^ 2
\end {equation}
which defines the tachyon mass.
This means that the tachyon mass obeys the relation:

\begin {equation}
m ^ 2 = \left( { 1 - D \over 12 } \right)^ 2
\end {equation}
As a check, we find that this simply
reproduces the usual relationship between the tachyon
mass and dimension. So therefore
the tachyon is massless in $D=1$ (or in two dimensions, if we
consider the Liouville field to be a dimension).

On the other hand, we can solve the mass-shell condition for
$\epsilon$ directly, yielding:
\begin {equation}
\epsilon = { - Q \pm \sqrt { Q ^ 2 - 8 + 4 p _ i ^ 2 } \over 2 }
\end {equation}
We shall be mainly interested in the case of two dimensions,
or $D = 1$, so we find $Q = 2 \sqrt 2$
and:
\begin {equation}
\epsilon = - \sqrt { 2} + \chi p
\end {equation}
where $\chi = \pm 1$ is called the \lq\lq chirality" of the tachyon
state.
The ground state, with arbitrary ghost number $\lambda$, can
therefore be written as:

\begin {equation}
| p , \epsilon , \lambda \rangle =
e ^ { i p X + \epsilon \phi + \lambda \sigma } ( 0 )
| 0 \rangle
\end {equation}
where $\sigma _ 0 | p , \lambda \rangle =
\lambda | p , \lambda \rangle$.
We will choose $\lambda =1 $ for the ghost vacuum.

Our tachyon state is then defined as
$ | p ^ M \rangle = | p , \epsilon , \lambda \rangle$.

In addition, we also have the $b-c$ ghost system.
We define the $SL(2,R)$ vacuum in the usual way:

\begin {equation}
\langle 0 | c _ { - 1 } c _ 0 c _ 1 | 0 \rangle = 0
\end {equation}
so the ghost system has background charge $-3$.
Then the ghost part of the tachyon field is given by
$ c _ 1 \bar c _ 1 | 0 \rangle$.

If we let $c _ 1 | 0 \rangle = | - \rangle$,
with ghost number $- 1 /2$,
then
the open string wave function is based on the
vacua $ | - \rangle$ and $c _ 0 | - \rangle = | + \rangle $.
For the closed string case, the string wave function
$|\Phi \rangle$ is based on four possible vacua,
so that:
\begin {equation}
| \Phi \rangle =
\varphi _ { -- } | - \rangle | - \rangle
+ \varphi _ { - + } |  - \rangle | + \rangle
+ \varphi _ { + - } | + \rangle
| - \rangle
+
\varphi _ { ++ } |+ \rangle |+ \rangle
\end {equation}

With this ground state, we can then construct the
vertex functions, once we know the Neumann functions. These
can be defined via the Green's function
on the string world sheet in the usual way:

\begin {eqnarray}
\label{eq:N}
N ( \rho _ r , \tilde \rho _ s ) & = &
- \delta _ { rs }
\Bigg \{
\sum _ { n \geq 1 } { 2 \over n }
e ^ { - n | \xi _ r - \tilde \xi _ s | }
\cos ( n \sigma _ r )
\cos ( n \tilde \sigma _ s )
- 2 {\rm max }
( \xi _ r , \tilde \xi _ s )
\Bigg \}
\nonumber \\
& + &
2 \sum _ { n,m \geq 0 } N _ { nm } ^ { rs }
e ^ { n \xi _ r + m \tilde \xi _ s }
\cos ( n \sigma _ r ) \cos ( n \tilde \sigma _ s )
\nonumber\\
& = &
\log | z - \tilde z |
+
\log | z - \tilde z ^ * |
\end {eqnarray}

By taking the Fourier transform of the previous equation, one
can invert the relation and find an expression for $N _ { nm} ^ { rs }$:

\begin {eqnarray}
N _ { nm } ^ { rs }
& = &
{ 1 \over nm }
\oint _ { z _ r } { dz  \over 2 \pi i }
\oint _ { z _ s } { d \tilde z \over 2 \pi i }
{ 1 \over ( z - \tilde z ) ^ 2 }
e ^ { - n \rho _ r ( z ) - m \tilde \rho _ s ( \tilde z ) }
\nonumber \\
N _ { n0 } ^ { rs } & = &
{ 1 \over n }
\oint _ { z _ r }
{ d z \over 2\pi i }
{ 1 \over z - z _ s }
e ^ { - n \rho _ r ( z ) }
\end {eqnarray}

In addition to these Neumann functions, we must also define
the $B_ N$ line integrals, which are found in the calculation of
any $N$-point tree graph and hence must appear in the vertex
function as well.

We have:

\begin {equation}
B _ N = \prod _ { j = 1 } ^ N
( b _ 0 - \bar b _ 0 ) _ j
\prod _ { k = 1 } ^ { 2N-6}  b _ { \mu _ k } d \tau _ k
\end {equation}
where:
\begin {equation}
b _ { \mu _ k } = \int { d ^ 2 \xi \over 2 \pi }
\left ( \mu _ k b ( z ) + {\rm c.c.} \right )
\end {equation}
where $\tau _ k$ are the modular parameters which specify
the polyhedra,
where $\mu _ k$ are the $2N-6$ Beltrami differentials which
correspond to the $2N-6$ quasi-conformal deformations which
typify how the polyhedral vertex function changes as the
moduli parameters $\tau _ i$ vary. These $\tau _ i$, in turn,
are functions of the angles $\theta _ { ij }$.

With these Neumann functions, we can construct the four-point
scattering amplitude. However, the Jacobian from the world sheet
to the complex $z$-plane requires elliptic integrals.

Our conventions are those of ref. [24].
First elliptic integrals are defined as:

\begin {eqnarray}
\label{eq:F}
F ( \phi , k ) & = &
\int _ 0 ^ y { dt \over
\sqrt { ( 1 - t ^ 2 ) ( 1 - k ^ 2 t ^ 2 ) } }
\nonumber \\
& = & \int _ 0 ^ \phi
{ d \theta
\over \sqrt {  ( 1 - k ^ 2 \sin ^ 2 \theta ) } }
\nonumber \\
& = &
{\rm sn } ^ { -1 } ( y , k )
\end {eqnarray}
where $y =\sin \phi$ and $\phi$ = am $u _ 1 $.

Second elliptic integrals are defined as:
\begin {eqnarray}
E ( \phi , k ) & = &
\int _ 0 ^ y { \sqrt { 1 - k ^ 2 t ^ 2 }
\over \sqrt { 1 - t ^ 2 } } dt
\nonumber \\
& = &
\int _ 0 ^ \phi \sqrt { 1 - k ^ 2 \sin ^ 2 \theta } d \theta
\end {eqnarray}

Third elliptic integrals are defined as:
\begin {eqnarray}
\label{eq:pi}
\Pi ( \phi , \alpha ^ 2 , k ) & = &
\int _ 0 ^  y
{ dt \over
( 1 - \alpha ^ 2 t ^ 2 ) \sqrt { ( 1 -t ^ 2 )
( 1 - k ^ 2 t ^ 2 ) } }
\nonumber\\
& = &
\int _ 0 ^ \phi
{ d \theta \over
( 1 - \alpha ^ 2 \sin ^ 2 \theta )
\sqrt { 1 - k ^ 2 \sin ^ 2 \theta } }
\nonumber \\
& = &
\int _ 0 ^ { u _ 1 }
{  d u \over 1 - \alpha ^ 2 {\rm sn } ^ 2 u }
\end {eqnarray}

Complete first elliptic integrals are defined as:
\begin {equation}
K( K ) = K = \int _ 0 ^ { \pi / 2 }
{ d \theta \over
\sqrt { 1 - k ^ 2 \sin ^ 2 \theta } }
= F ( \pi / 2 , k )
\end {equation}

Complete second elliptic integrals are defined as:
\begin {equation}
E ( \pi / 2 , k ) = E =
\int _ 0 ^ { \pi / 2 }
\sqrt { 1 -k ^ 2 \sin ^ 2\theta }
d \theta
\end {equation}

Heuman's lambda function is defined as:
\begin {equation}
\label{eq:lambda}
\Lambda _ 0 ( \phi, k ) =
{ 2 \over \pi }
[ EF (\phi , k ' ) +
K E ( \phi , k ' )
- KF ( \pi , k ' )]
\end {equation}

The Jacobi zeta function is defined as:
\begin {equation}
\label{eq:Z}
Z ( \phi , k ) = E ( \phi , k )
- { E \over K } F ( \phi , k )
\end {equation}

In the text, we have used the following differential
equations:

\begin {eqnarray}
\label{eq:deriv}
{ d \over d k }
\left [ K ( k ' ) Z ( \beta _ i , k ' ) \right ]
& = &
{ k ' E ( K ' ) \over k ^ 2 }
{ \sin \beta _ i \cos \beta _ i \over
r (\beta _ i , k ' ) }
\nonumber \\
{ d \over d k }
\Lambda _ 0 & = &
{ 2 \over \pi k }
\left [ E ( k ) - K ( k ) \right ]
{\sin \beta _ i \cos  \beta _ i
\over r ( \beta _ i , k ' ) }
\nonumber \\
{ d \over d \beta _ i }
\left [ K ( k ' )  Z ( \beta _ i , k ' )
\right ]
& = &
{ r ^ 2 ( \beta _ i , k ' ) K ( k ' ) -
E ( k ' )
\over
r ( \beta _ i , k ' )
}
\nonumber \\
{ d \over d \beta _ i }
\Lambda _ 0 ( \beta _ i , k ) & = &
{ 2 \over \pi r ( \beta _ i , k ' ) }
\left [ E( k ) -
k ^ { 2 \prime }
\sin ^ 2 \beta _ i  K ( k ) \right ]
\end {eqnarray}

\section {References}

\noindent 1.
For reviews, see refs. 1-3:
\lq\lq Lectures on 2D Gravity and 2D String Theory,"
by P. Ginsparg and G. Moore, YCTP-P23-92.

\noindent 2. \lq\lq Developments in 2D String Theory,"
by A. Jevicki, BROWN-HET-918.

\noindent 3. I.R. Klebanov, \lq\lq String Theory in Two Dimensions,"
Proceedings of the Trieste Spring School 1991, eds. J. Harvey et. al.
World Scientific, Singapore, 1992.

\noindent 4. D.J. Gross, I.R. Klebanov, and
M.J. Newman, {\it Nucl. Phys.} {\bf B350}, 621 (1991).

\noindent 5. D.J. Gross and I.R. Klebanov,
{\it Nucl. Phys.} {\bf B352} 671 (1991).

\noindent 6. A.M. Polyakov, {\it Mod. Phys. Lett.}
{\bf A6}, 635 (1991).

\noindent 7. I.R. Klebanov and A.M. Polyakov,
{\it Mod. Phys. Lett.} {\bf A6}, 3273, 1991;

\noindent 8. A. M. Polyakov, {\it Phys. Lett.} {\bf B103}, 207, 211, (1981).

\noindent 9. \lq\lq Lecture Notes on 2D Quantum Gravity and
Liouville Theory," E. D'Hoker, UCLA/91/TEP/35.

\noindent 10.  I. Ya. Aref' eva, P.B. Medvedev, and A.P.
Zubarev, Steklov Inst. preprint.

C.R. Preitschopf and C.B. Thorn, UFIFT-HEP-90-17.

\noindent 11. B. Urosevic, BROWN-HET-899.

\noindent 12.  M. Kaku, in {\it Functional
Integration, Geometry, and Strings},
Birkhauser Press, Berlin, (1989).

M. Kaku, {\it Phys. Rev.} {\bf D41}, 3734 (1990).

\noindent 13. T. Kugo, H. Kunitomo, and K. Suehiro, {\it
Phys. Lett.} {\bf
226B}, 48 (1989).

\noindent 14. M. Saadi and B. Zwiebach, {\it Ann. Phys.} {\bf 192}, 213 (1989).

\noindent 15. M. Kaku, {\it Phys. Lett.} {\bf 250B}, 64 (1990).

\noindent 16. B. Zwiebach, {\it Nucl. Phys.} {\bf B390}, 33 (1993).

\noindent 17. S. Mandelstam, Nucl. Phys.
{\bf B69}, 77 (1974).

\noindent 18. H. Hata, K. Itoh, T. Kugo,
H. Kunitomo and K. Ogawa, Phys. Rev. {\bf D34},
2360 (1986).

\noindent 19.
K. Suehiro, {\it Nucl. Phys.} {\bf B296}, 333 (1988).

\noindent 20. M. Kaku and J. Lykken, {\it Phys. Rev.}
{\bf D38}, 3052 (1988).

\noindent 21. E. Witten, {\it Nucl. Phys.}
{\bf B268}, 253 (1986).

\noindent 22. S. Giddings, {\it Nucl. Phys.} {\bf B278},
242 (1986).

\noindent 23. M. Kaku and L. Hua,
{\it Phys. Rev.} {\bf D41}, 3748 (1990).

\noindent 24. P.F. Bryd and M.D. Friedman, {\it Handbook of
Elliptic Integrals for Engineers and Scientists},
Springer, New York (1971).

\noindent 25. S.R. Das and A. Jevicki, {\it Mod. Phys. Lett.}
{\bf A5}, 1639 (1990).

\noindent 26. A. Jevicki and B. Sakita, {\it Nucl. Phys.}
{\bf B165}, 511 (1980).

\end {document}